# Investigating the difference in mechanical stability of retained austenite in bainitic and martensitic high-carbon bearing steels using in situ neutron diffraction and crystal plasticity modeling

Rohit Voothaluru[a,*], Vikram Bedekar[a], Dunji Yu[b], Qingge Xie[b], Ke An[b], R. Scott Hyde[a]

[a]The Timken Company, North Canton, OH, 44720 USA

[b]Chemical and Engineering Materials Division, Spallation Neutron Source, Oak Ridge National Laboratory, Oak Ridge, TN 37831 USA

*Corresponding Author: Email Address: rohit.voothaluru@timken.com. Tel. +1-(234)-262-3984

**Abstract**

In situ neutron diffraction of the uniaxial tension test was used to study the effect of the surrounding matrix microstructure on the mechanical stability of retained austenite in high-carbon bearing steels. Comparing the samples with bainitic microstructures to those with martensitic ones it was found that the retained austenite in a bainitic matrix starts transforming into martensite at a lower strain compared to that within a martensitic matrix. On the other hand, the rate of transformation of the austenite was found to be higher within a martensitic microstructure. Crystal plasticity modeling was used to analyze the transformation phenomenon in these two microstructures and determine the effect of surrounding microstructure on elastic, plastic and transformation components of the strain. The results showed that the predominant difference in the deformation accumulated was from the transformation strain and the critical transformation driving force within the two microstructures. The retained austenite was more stable for identical loading conditions in case of martensitic matrix compared to the bainitic one. It was also observed that the initial volume fraction of retained austenite within the bainitic matrix would alter the onset of transformation to martensite but not the rate of transformation.

# 1. Introduction

Today's industrial applications involve bearings and gears that sustain severe static and dynamic loads while serving reliably in extreme environments[1]. High-carbon and chromium-containing steels are widely used in the current bearing industry, where the steel is often heat-treated to obtain either martensitic or bainitic microstructures [2]. High-carbon martensitic steels are very well known for their high strength and hardness properties and have been used extensively for bearings over the past century [1,3]. Bainitic or martensitic microstructures are often obtained together with a retained volume fraction of austenite. The volume fraction of the retained austenite (RA) and its morphology are of great interest in the applications of rolling element bearings [4].

The retained austenite in bearing steels is predominantly present in either film-like form or block-type form [5,6]. The films of austenite which are less stable when it comes to decomposition during tempering are however, well-known to be very stable under the application of stress [7]. The film-like austenite has been reported to be very stable due to its carbon enrichment, and such fine microstructures have been known to give an excellent combination of strength and toughness [8]. One of the roles of retained austenite in engineering applications is the enhancement of the ductility of steel which has been reported by De Cooman [9] and Jacques [10]. The strain induced transformation of retained austenite increases the work-hardening rate sufficiently to delay plastic instabilities in application loading conditions. Over the years, several studies have found that the presence of retained austenite films around the martensite helps improve the fracture toughness and fatigue of steels in bearings and gears [11–13]. The improved fatigue resistance in the presence of RA has been attributed to the enhanced mean compressive stress created by the transformation of the RA into martensite [14]. Voskamp [15] found that the transformation of RA depended on the cumulative effect of cyclic loading, and that the transformation of RA is very sensitive during the entire life cycle of a bearing [16]. It has been well established by now that the kinematic stability of RA is influenced by several factors including size, morphology, and %C in the austenite and surrounding microstructure [17–19]. Recently, Xiong *et al.* [20] found that low-carbon, film-like austenite is more stable than high-carbon blocky austenite. Ryu *et al.* [18] found that the stability of retained austenite depends upon the strain partitioning with the surrounding matrix. In their study using steels with different Al content, the authors found that the grain size distribution of the ferritic regions also plays an influential role in the stability of retained austenite. The authors [15] caution that chemical composition or strain partitioning should not be the sole factor considered in the determination of RA stability. The determination of RA stability actually is extremely challenging due to many interdependent factors. The majority of the existing research on the stability of austenite and transformation-induced plasticity has been focused on transformation-induced plasticity (TRIP) steels [21–24]. Jacques *et al.* [17] studied the mechanical stability of RA in TRIP steels and found that the harder martensite shielded the RA. Hidalgo *et al.* [25] studied austenite/martensite microstructures by varying the degrees of tempering and found that the surrounding microstructure influences the stability of RA. But there has been little progress in the case of martensitic or bainitic (58-60 HRc) bearing steels.

Recent studies on through-hardened bearing steels using X-ray synchrotron [21] and neutron diffraction [23,24,26,27] have shown considerable promise for understanding the behavior of retained austenite in a quantitative and discrete manner. With the advances afforded by the use of neutron diffraction, real-time and continuous studies can be performed in a precise and accurate manner. However, in spite of the significant progress in understanding the deformation mechanisms and stability of austenite in steels, there is very little information regarding the effect of the surrounding microstructure on the onset and rate of austenite transformation in through-hardened bearing steels. The effect of the surrounding microstructure on the mechanical stability of retained austenite has not yet been comprehensively studied. Also, there has been limited research on the comparison between the mechanical stability of austenite within martensitic and within bainitic microstructures. There is also very little information in literature on the actual stress state and evolution of the stress state of the austenite when surrounded by bainite and martensite.

In order to address this gap and further provide a quantitative estimation of the mechanical stability of retained austenite in through-hardened bearing steels, in this work the authors studied the kinematic stability of retained austenite in through-hardened ASTM A485 Grade 1 steel under uniaxial tensile loading. The effects of varying soak times resulting in different amounts of initial RA% within the bainitic microstructure was studied using in-situ neutron diffraction. The effect of bainite and martensite in the surrounding microstructure was also studied to compare the effect of the surrounding microstructure on the mechanical stability of the retained austenite. The results were then analyzed using a hybrid polycrystal plasticity finite element model to determine the elastic, plastic and transformation plasticity components of the deformation so as to understand the predominant portion of the deformation causing the differences in the relative mechanical stability of the retained austenite.

## 2. Materials and Methods

### 2.1 Material Characterization and Sample Preparation

The samples used in this study were dogbone tensile specimens of A485 Grade 1 (A485-1) steel, the composition of which is shown in Table 1. Three samples with different surrounding microstructures (two bainitic, B1 and B2 and one martensitic, M1) were created by altering the heat treatment. Sample B1, with a bainitic microstructure containing 18% retained austenite, was prepared by austenitizing at 850°C for 45 minutes followed by soaking in a salt bath at 230°C for 2 hours. Sample B2 was austenitized at the same temperature and time, but soaked at 230°C for 4 hours. The longer hold time allowed continued transformation of the austenite, reducing its volume to 9.1%. It should be noted that the longer soak times transform the less stable austenite into bainite, leaving behind austenite that is stable and enriched with carbon. Sample M1, with a martensitic microstructure and 18% retained austenite was prepared by hardening at 850°C followed by tempering at 180°C for 1.3 hours.

**Table 1:** Chemical Composition of A485-1 Steel (wt.%) [26]

| C  | Mn    | Si   | Cr    | Ni    | P      | S      | Fe  |
|----|-------|------|-------|-------|--------|--------|-----|
| 1% | 1.09% | 0.6% | 1.06% | 0.11% | 0.013% | 0.012% | Bal |

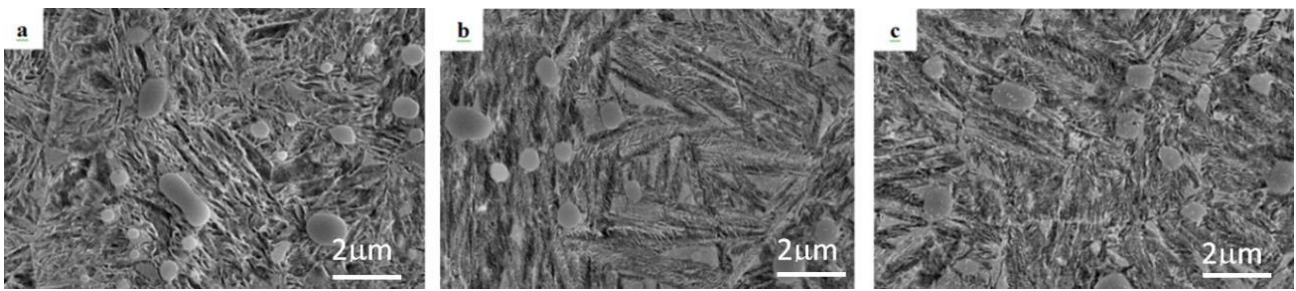

**Fig. 1:** SEM Images of: a) M1 (Martensitic) [26]; b) B1 (Bainitic with 2-Hour Soak); and c) B2 (Bainitic with 4-Hour Soak) Microstructures

The %C in the austenite was calculated using a technique described by Jatczak *et al.*[28] and was found to be higher in the bainitic samples B1 (0.98%) and B2 (1.09%) than in the martensitic M1 sample (0.9%). The hardness and microstructural parameters estimated for the three specimens are summarized in Table 2. The specimen dimensions were chosen to allow appropriate testing for neutron diffraction and the manufacturing process was carefully monitored to ensure that the surrounding martensitic and bainitic microstructures could be compared appropriately. SEM microstructure images showed a mix of film-like and blocky RA structures in the bainitic sample B1, as shown in Fig. 1. The retained austenite in sample B2 was less blocky and contained more film-like austenite compared to samples M1 and B1. The martensitic M1 sample has a mix of blocky and fine retained austenite.

**2.2 In-situ Neutron Diffraction**

The in-situ neutron diffraction experiment was conducted during monotonic tension testing of the three specimens as depicted by the schematic in the Fig. 2. Understanding the deformation modes and quantifying the effect of surrounding microstructure on individual phases and the mode of deformation can be realized using neutrons. The large penetration depth of the neutrons and the volume-averaged nature of the bulk measurement which is characteristic of a scattering beam allow the determination of the localized deformation at the lattice level in these different specimens [28]. The in-situ neutron diffraction experiments were conducted on the VULCAN engineering diffractometer at the Spallation Neutron Source in the Oak Ridge National Laboratory. The VULCAN time of flight diffractometer enabled rapid collection of structural changes in the sample under dynamic loading conditions. The uniaxial tension test was performed at a strain rate of $5\times10^{-6}$. The tensile specimen was exposed to a 5mm x 5mm neutron beam with a 5mm collimator, allowing data collection over a 125mm$^3$ gauge volume. The data was collected in situ using two detector banks located in the longitudinal and transverse directions. Data was chopped at 2-minute intervals and analyzed using VDRIVE (Vulcan Data Reduction and Interactive Visualization) software [29]. Reference scans were conducted in unloaded condition over a period of ten minutes to minimize the $d_0$ propagated statistical error [30]. The peak positions were fit using single peak fit.

**Table 2:** Specimen Details for Martensitic and Bainitic Samples

| Sample ID | Hardness (HRc) | %RA | %C in RA |
|---|---|---|---|
| B1 | 58.6 | 17.7 | 0.98 |
| B2 | 58.8 | 9.1 | 1.09 |
| M1 | 61.0 | 18.0 | 0.9 |

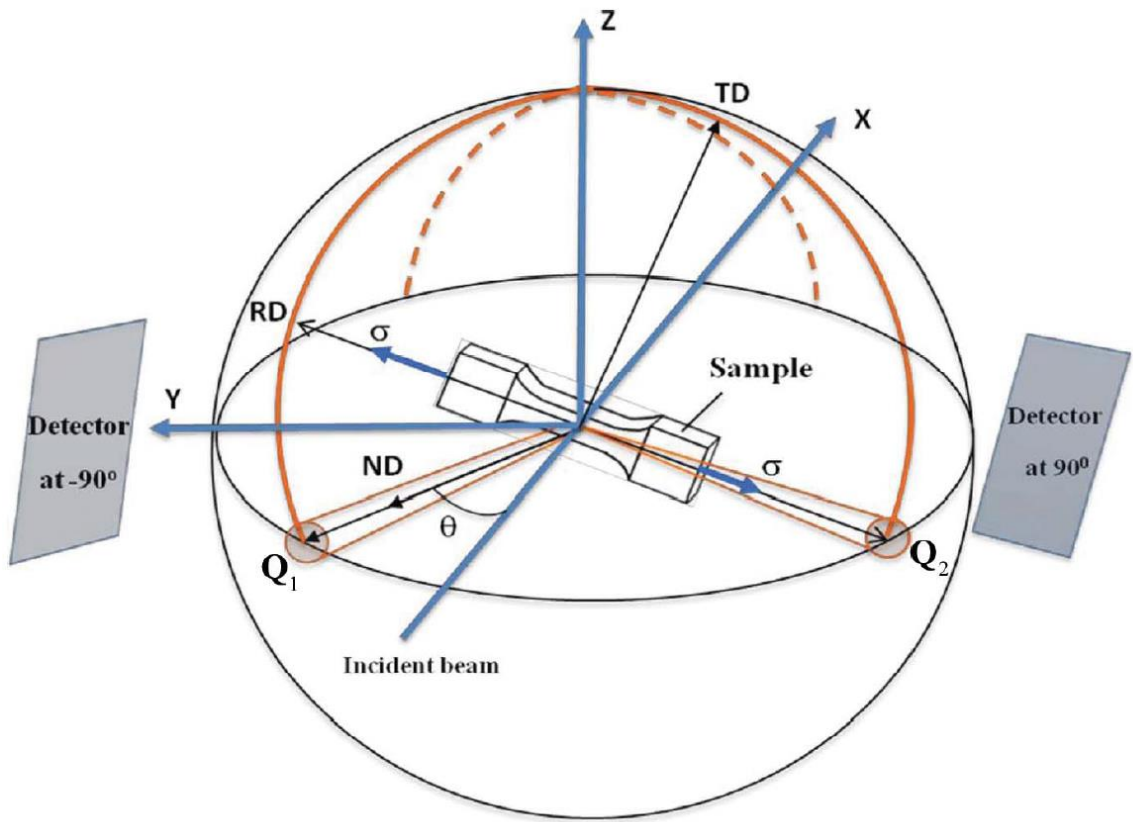

**Fig. 2**: Schematic for the in situ measurements (loading) at the VULCAN diffractometer: XYZ is the instrument coordinate system, xyz is the sample coordinate system (identified, respectively, with the main processing directions, rolling, RD, transverse, TD, and normal, ND), σ is the mechanical loading along the RD, θ is the diffraction angle, and Q1 and Q2 are the scattering vectors [31]

## 3. Constitutive Model

From a purely mechanistic analysis, the rates of transformation and the variations in the quantified deformation during uniaxial tension testing of the test specimens can be attributed to several factors because the deformation in the microstructure consists of elastic, plastic and transformation plasticity components. In order to analyze the neutron diffraction data and understand the underlying deformation mechanisms within the two surrounding microstructures studied in this work, the empirical data must be coupled with a suitable computational or analytical technique that will help to determine the major factor contributing to the differences observed in the mechanical characterization. In order to address this we used a hybrid polycrystal plasticity finite element (CPFE) approach to study the differences between the three specimens.

The use of a crystal plasticity formulation to describe the transformation plasticity and deformation of bearing steels has demonstrated significant success with martensitic through-hardened and case-carburized microstructures. Recent works by Woo *et al.*, Kim *et al.* and Voothaluru *et al.*[23,24,26] have shown that polycrystal plasticity finite element modeling can help quantify the micro-mechanical parameters, as it captures the deformation mechanisms that occur during the in situ neutron diffraction of multi-phase steels. For this work, the CPFE model was developed in ABAQUS based upon the implementation of the constitutive model proposed by Turteltaub and Suiker [32]. The implementation of the constitutive model into a user material subroutine (UMAT) was carried out by following the methods of Voothaluru and his co-authors[33]. The computational model is semi-empirical in nature, and the material parameters are evaluated for elastic, plastic and transformation plasticity components by determining the best fit of the simulated result with the empirical data. For the CPFE model, the total deformation gradient $\boldsymbol{F}$ is given by:

$$\boldsymbol{F} = \boldsymbol{F}^e . \boldsymbol{F}^p . \boldsymbol{F}^{tr} \qquad (2)$$

where $\boldsymbol{F}^{tr}$ is the transformation gradient that accounts for the volumetric strain produced by austenite-martensite phase transformation, $\boldsymbol{F}^p$ accounts for the polycrystalline plasticity and $\boldsymbol{F}^e$ is the elastic deformation gradient.

In this work, the model assumes that the 48 BCC slip systems within a representative volume element (RVE) will exhibit behavior approximately similar to that of tempered martensite and bainite. Both the bainitic and martensitic microstructures are modeled as BCC for computational purposes, as discussed in prior works [34–37] The shear and transformation rates are computed at the end of each time step. The plastic shearing rate $\dot{\gamma}^{(\alpha)}$ on the $\alpha^{th}$ slip system is governed by the rate-dependent flow rule:

$$\dot{\gamma}^{(\alpha)} = \dot{\gamma}_0 \cdot \left|\frac{\tau^{(\alpha)} - \chi^\alpha}{g^{(\alpha)}}\right|^m \cdot sgn(\tau^{(\alpha)} - \chi^{(\alpha)}) \qquad (3)$$

where $m$ is the strain rate sensitivity exponent, $\dot{\gamma}_0$ is the shearing rate coefficient, $g^{(\alpha)}$ is the drag stress, $\chi^{(\alpha)}$ is the back stress and $\tau^{(\alpha)}$ is the resolved shear stress on the $\alpha^{th}$ slip system. The resolved shear stress on each slip system is related to the Cauchy stress tensor $\boldsymbol{\sigma}$, according to:

$$\tau^\alpha = \boldsymbol{\sigma} : (\boldsymbol{s}^{(\alpha)} \otimes \boldsymbol{m}^{(\alpha)}) \qquad (4)$$

Here, $\boldsymbol{s}^{(\alpha)}$ and $\boldsymbol{m}^{(\alpha)}$ are unit vectors defined in the crystal coordinate system representing the slip direction and the slip plane normal respectively. The drag stress and back stress evolution follow the expressions in Eq. (5) and (6), respectively:

$$\dot{g}^{(\alpha)} = \sum_{\beta=1}^{N_{slip}} H_{dir} \cdot |\dot{\gamma}^{(\beta)}| - g^{(\alpha)} \cdot \sum_{\beta=1}^{N_{slip}} H_{dyn} \cdot |\dot{\gamma}^{(\beta)}| \qquad (5)$$

$$\dot{\chi}^{(\alpha)} = A_{dir} \cdot \dot{\gamma}^{(\alpha)} - \chi^{(\alpha)} \cdot A_{dyn} \cdot |\dot{\gamma}^{(\alpha)}| \qquad (6)$$

where $H_{dir}$ is the isotropic hardening coefficient, $H_{dyn}$ is the dynamic recovery coefficient and $A_{dyn}$ is the dynamic recovery coefficient for the back stress. The rate of volume fraction transformation $\dot{\xi}^{(\lambda)}$ on a transformation system $\lambda$ is given by:

$$\dot{\xi}^{(\lambda)} = \dot{\xi}_{max} \cdot \tanh\left(\frac{1}{v^{tr}} \cdot \left(\frac{<f_{tr}^\lambda - f_{cr}^\lambda>}{f_{cr}^\lambda}\right)\right) \qquad (7)$$

where $\dot{\xi}_{max}$ is the maximum rate of transformation, $v^{tr}$ is the viscosity parameter, $f_{cr}^\lambda$ is the critical driving stress and $f_{tr}^\lambda$ is the driving stress on the $\lambda^{th}$ transformation system. The driving stress on the $\lambda^{th}$ transformation system is related to the transformation and habit vectors $\hat{b}^\lambda$ and $\hat{n}^\lambda$ and the Cauchy stress tensor by:

$$f_{tr}^\lambda = \boldsymbol{\sigma}:(\gamma_T.\hat{\boldsymbol{b}}^\lambda \otimes \hat{\boldsymbol{n}}^\lambda) \tag{8}$$

where $\gamma_T$ is the shape strain magnitude — a parameter that is uniform for all transformation systems. The rate of change of the volume fraction transformed from austenite to martensite is given by Eq. (9), where $\dot{V}_{trans}$ is the rate of transformation of retained austenite.

$$\dot{V}_{trans} = \sum_{\lambda=1}^{N_{trans}} \dot{\xi}^{(\lambda)} \tag{9}$$

The critical driving force $f_{cr}^\lambda$ is controlled by the transformation rates, as shown in Eq. (10). Here, $Q$ is the transformation hardening coefficient along transformation plane $\eta$. $Q$ accounts for the increased resistance to transformation as more of the retained austenite becomes surrounded by transformed martensite. The increased resistance to transformation in any system is assumed to be the same [35].

$$\dot{f}_{cr}^\lambda = \sum_{\eta=1}^{N_{trans}} Q.|\dot{\xi}^{(\eta)}| \tag{10}$$

The transformation gradient that accounts for the volumetric transformation from austenite to martensite is given by:

$$\boldsymbol{F}^{tr} = \sum_{\lambda=1}^{N} \gamma_T \xi^{(\alpha)} \hat{\boldsymbol{b}}^\alpha \otimes \boldsymbol{m}^{(\alpha)} \tag{11}$$

The model described here was used to study the micro-mechanical response of the RVE in order to determine the single crystal elastic constants, micro-plasticity parameters and transformation plasticity parameters of the hybrid representation. This determination allows a comparison between the three specimens and provides a means for understanding the influence of the microstructure on the variation in austenite stability.

## 4. Results
### 4.1 Experiment Results

The macroscopic stress-strain response, RA transformation for the bainitic samples B1 and B2, and along with results for the martensitic sample M1 from a previous study [26] are shown in Fig. 3. In sample B1 (bainitic microstructure), the apparent macroscopic yield point was observed to be at a true stress of 722 MPa (0.344% strain). In sample B2, which had lower retained austenite (9%) in a bainitic microstructure, the deviation from linearity occurred at a true stress of 1340 MPa (0.7% strain). Rietveld analysis and single peak fitting were performed on the diffraction data using the General Structural Analysis System (GSAS) software package, and the volume fraction of RA was estimated for the duration of the uniaxial tension test. From Fig. 3, we can see that the onset of RA transformation into product martensite for each specimen, B1, B2 and M1, coincides with its corresponding macroscopic yield point from the true stress and true strain curve. This can be attributed to the fact that under uniaxial tension, the specimens are more susceptible to undergoing an austenite transformation that is strain-induced or strain-assisted, as hypothesized and discussed in prior works in literature [31–33].

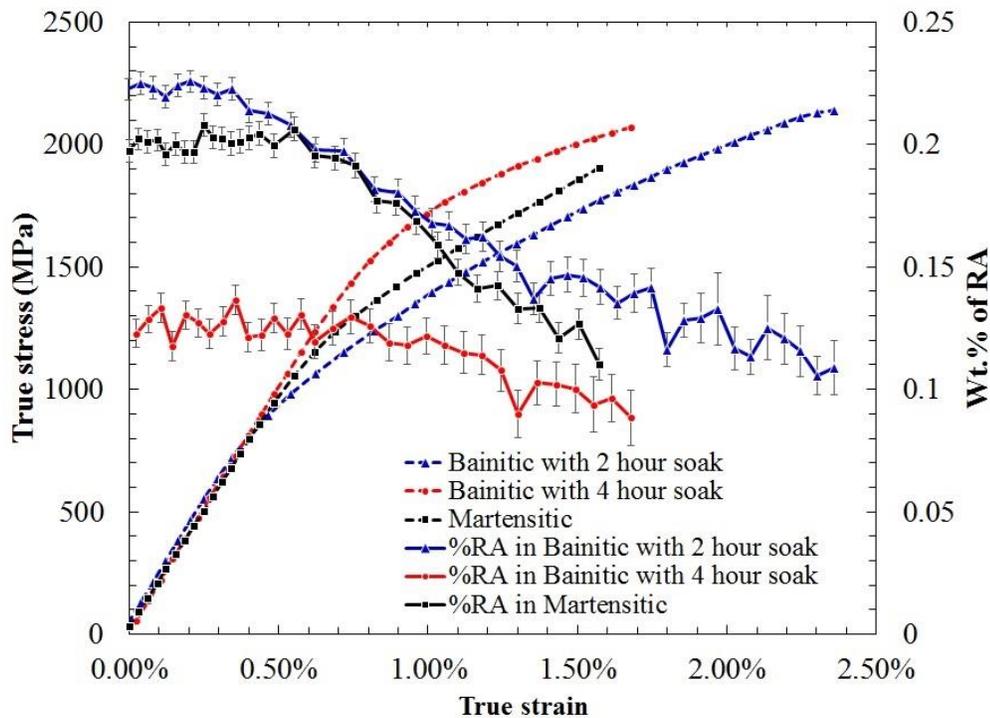

**Fig. 3:** Macroscopic Stress vs. True Strain with Respect to RA Transformation for B1, B2 and M1 Samples

As a result, the austenite transformation appears to initiate only after some amount of micro-plasticity is observed within the austenite, which can be tracked using the lattice strain estimated for the specimens using in situ neutron diffraction. Fig. 4 shows the lattice strain as a function of true stress, calculated using Eq. (1) using Rietveld refinement.

$$\varepsilon = \frac{d-d_0}{d_0} \quad (1)$$

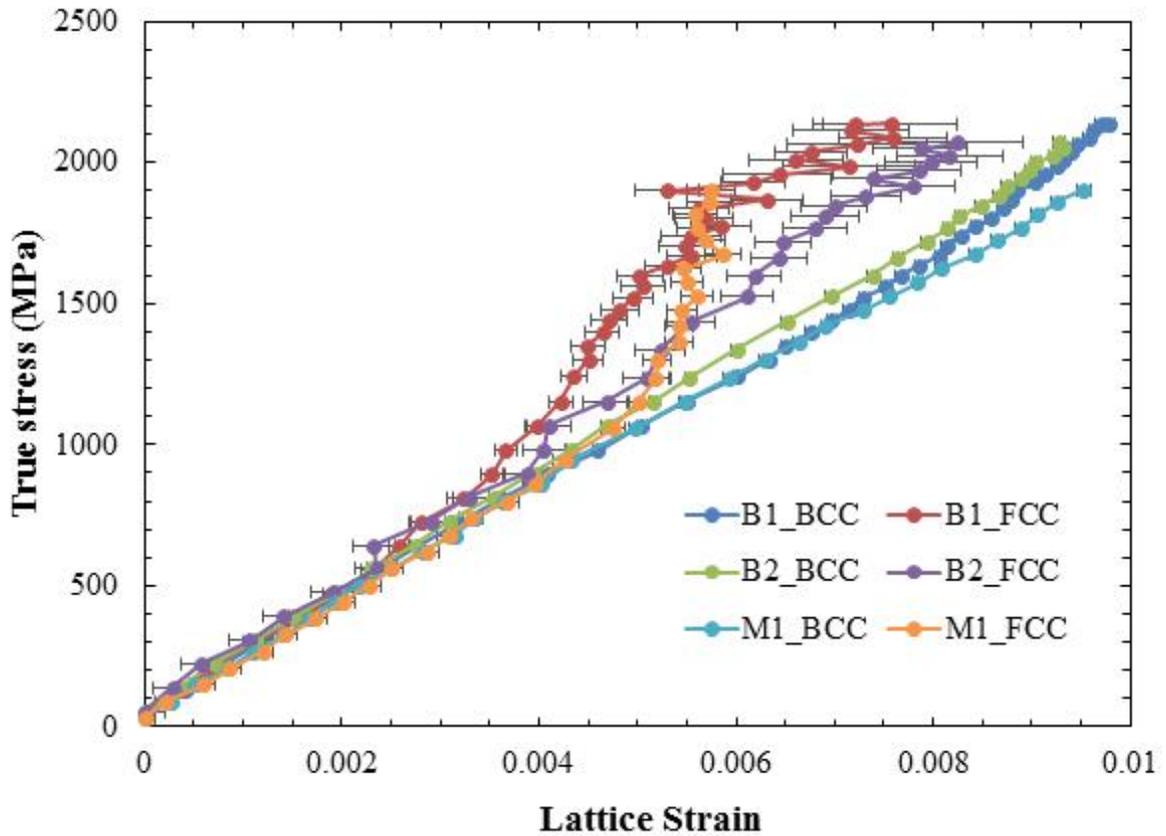

**Fig. 4:** Lattice Strain Plot for the Three Specimens Evaluated from In Situ Neutron Diffraction

The lattice strain data for the austenite (FCC) in all three specimens shows that the response is elastic until the macroscopic yield point is reached. The onset of RA transformation into martensite matches with the deviation from linearity of the lattice strain for austenite within all three materials. It can be seen that the TRIP transformation was triggered only when the plastic deformation occurred in the FCC austenite. In addition, it can be observed that the entire BCC matrix, which in this case represents both, the bainite and martensite (approximated from its BCT structure), remains perfectly elastic for the full loading period. Thus it can be concluded that the macroscopic yield point, onset of RA transformation, and deviation of the FCC lattice are all indications that the critical transformation driving force was being met.

From Fig. 3 and Fig. 4, it can be seen that the RA in B2 is more stable compared to B1. The longer soak time facilitated the continued transformation of RA into bainite. In this process, the more stable austenite with higher %C would be retained; as a result, the RA in B2 did not transform until a macroscopic stress of 1340 MPa (and a corresponding strain of 0.7%) was reached. The apparent higher mechanical stability of RA in B2 would be due to the lower volume fraction of blocky RA present in the microstructure. As discussed by Olson and Cohen [38], in the case of strain-induced austenite transformation, the presence of a lower volume fraction of RA results in a lower probability of product martensite nuclei being formed under similar loading conditions. It is also important to note that the presence of more film-like austenite in B2 compared to B1 would further cause the RA to be more stable under similar mechanical loading conditions.

In a similar fashion, comparing the results for bainitic and martensitic surrounding microstructures by looking at B1 and M1 we can see from Fig. 3 and Fig. 4 that the onset of transformation of the retained austenite into martensite occurred earlier in the bainitic microstructure than in the martensitic. It can be observed that under identical stress states, the austenite was relatively more stable under uniaxial tension loading in the martensitic microstructure, and needed 40% more true strain in order for RA transformation to initiate. This is intriguing, considering that the martensitic sample did not have film-like RA and the %C in the austenite was lower than that of both the bainitic samples. The higher relative stability of RA in the martensitic microstructure could, however, be attributed to the fact that the harder martensite appears to shield the austenite by accommodating more strain during loading as hypothesized and discussed in some earlier works by Jacques *et al.*[17] and Ryu *et al.*[18]. This shows that for same amount of initial RA, the martensitic sample would be more stable at comparable strains when compared to the bainitic sample.

On the other hand, from Fig. 3 it can also be seen that during the monotonic loading of both the B1 and M1 specimens, the martensitic specimen ended with 8.5% RA at fracture, whereas the bainitic specimen B1 had 11% RA at failure. This shows that while the austenite is more stable in the martensitic microstructure for specific applied loads, the rate of austenite transformation after the transformation has initiated is relatively greater in the martensitic specimen than in the bainitic one. After the onset of the austenite transformation, the blocky austenite in sample M1 transformed at a faster rate compared to the austenite in sample B1. In order to further understand the effect of the surrounding microstructure on the different deformation modes in these samples, the crystal plasticity model was used to determine the material parameters using a purely mechanistic analysis so as to understand the rate of transformation and the criterion for initial transformation.

**4.2 Crystal Plasticity Model Results**

In order to determine and quantify the differences in transformation of RA the constitutive model was implemented in ABAQUS using a UMAT as described earlier. The models for the B1, B2 and M1 specimens were created by generating an RVE with C3D8R elements in ABAQUS. The boundary conditions were chosen so as to simulate the real deformation as closely as possible. As a result, periodic boundary conditions were implemented in line with existing models [39,40] to accurately depict the real loading conditions on the RVE. A complete description of the boundary conditions, RVE generation [37] and UMAT implementation have been discussed in the prior work. The computational models generated were set up using 1000 elements (10x10x10) with initial volume fractions of retained austenite that matched the measured values from the microstructural characterization. The hybrid crystal plasticity formulation employed herein allowed us to generate identical RVEs with the orientation information populated through a Monte Carlo simulation. In addition, the material response from the finite element model was analyzed from 50 such instantiations of the RVE. The values of the shape strain magnitude ($\gamma_T$), transformation viscosity parameter ($v_{tr}$) and maximum rate of transformation ($\dot{\xi}_{max}$) in the austenite transformation model were assumed to be constant for this current composition of steel for all three specimens, and were set to 0.1809, 0.17 and 0.003 s$^{-1}$, respectively [41]. *H$_{dir}$* was set to 6.9GPa for the BCC phase of all three specimens and 6.4GPa for the FCC austenite in all three specimens [26]. In case of the FCC

Austenite $H_{dir}$ was set to 6.4GPa and the initial drag stress fit was set to 560MPa from Voothaluru et al. [26]. The retained austenite volume fraction in each of the three specimens was estimated using in situ neutron diffraction; the model was updated to reflect this in the input parameters. The material model and the corresponding input parameters were determined by performing a two-step iterative fit following the numerical implementation scheme proposed by Alley and Neu [36]. The single crystal elastic constants for the BCC and FCC phases were chosen to be constant across the three specimens as the material response under elastic loads was largely same as seen in Fig. 3 and Fig. 4. The single crystal elastic constants for the BCC and FCC phases are summarized in Table 3. The transformation plasticity parameters and microplasticity parameters were iteratively determined to fit the micromechanical response from the model to the empirical results from neutron diffraction. The material parameters used for the model are summarized in Table 4. The RVE was built in ABAQUS with C3D8R elements and periodic boundary conditions were applied to the RVE along the loading axis so as to simulate the representative loads from the uniaxial tension test. Fig. 5 illustrates the boundary conditions on the RVE. The individual phases within the microstructure were randomly flagged so as to distribute a volume fraction of austenite that is representative of the sample being modeled. Fig 6 shows an illustration of two such RVEs representing B1 and B2 samples.

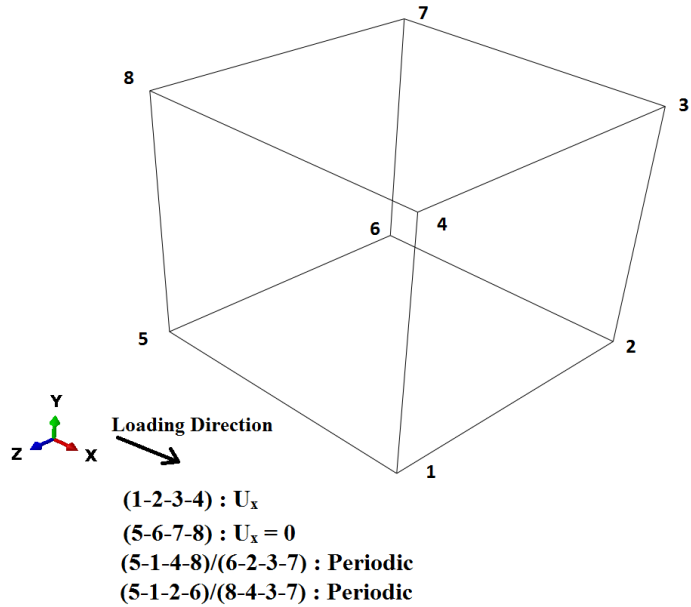

Fig. 5 Boundary Conditions for the CPFE modeling

**Table 3:** Material Model Parameters (GPa)

| Martensite/Bainite | | | Austenite | | |
|---|---|---|---|---|---|
| C11 | C12 | C44 | C11 | C12 | C14 |
| 278.7 | 114.2 | 90.2 | 229.1 | 101.2 | 85.4 |

**Table 4:** Transformation plasticity and microplasticity parameters

| Specimen | $g_0^{(\alpha)}$ (MPa) | $RA_{initial}$ | $f_{cr}^{\lambda}$ (MPa) | $Q$ (MPa) |
|---|---|---|---|---|
| **B1** | 690 | 18% | 80.3 | 640 |
| **B2** | 690 | 9% | 145 | 670 |
| **M1** | 840 | 18% | 97.1 | 549 |

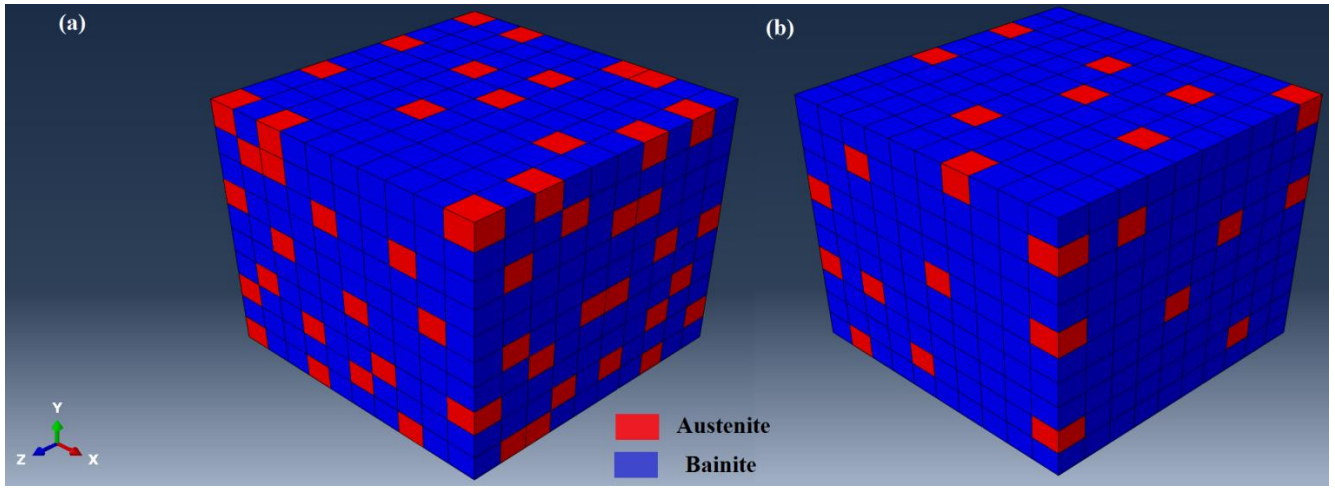

Fig. 6 (a) Illustration of an RVE instantiation for B1 Sample (b) Illustration of an RVE instantiation for B2 Sample in ABAQUS

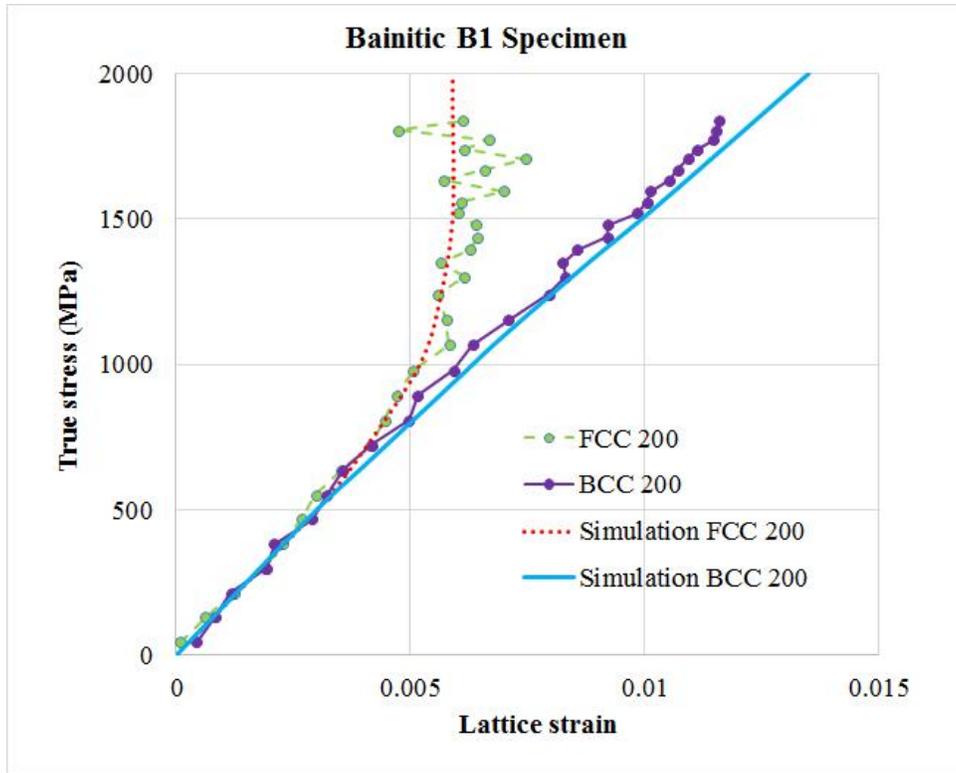

**Fig. 7:** CPFE Estimation and Experiment Results for lattice strains in Bainitic B1 Specimen

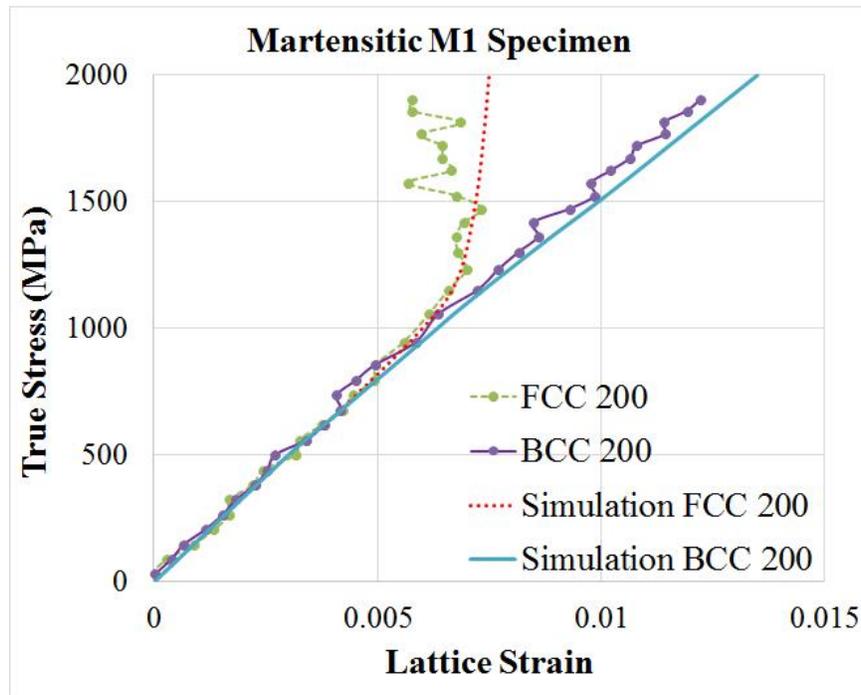

**Fig. 8:** CPFE Estimation and Experiment Results for lattice strains in Martensitic M1 Specimen [26]

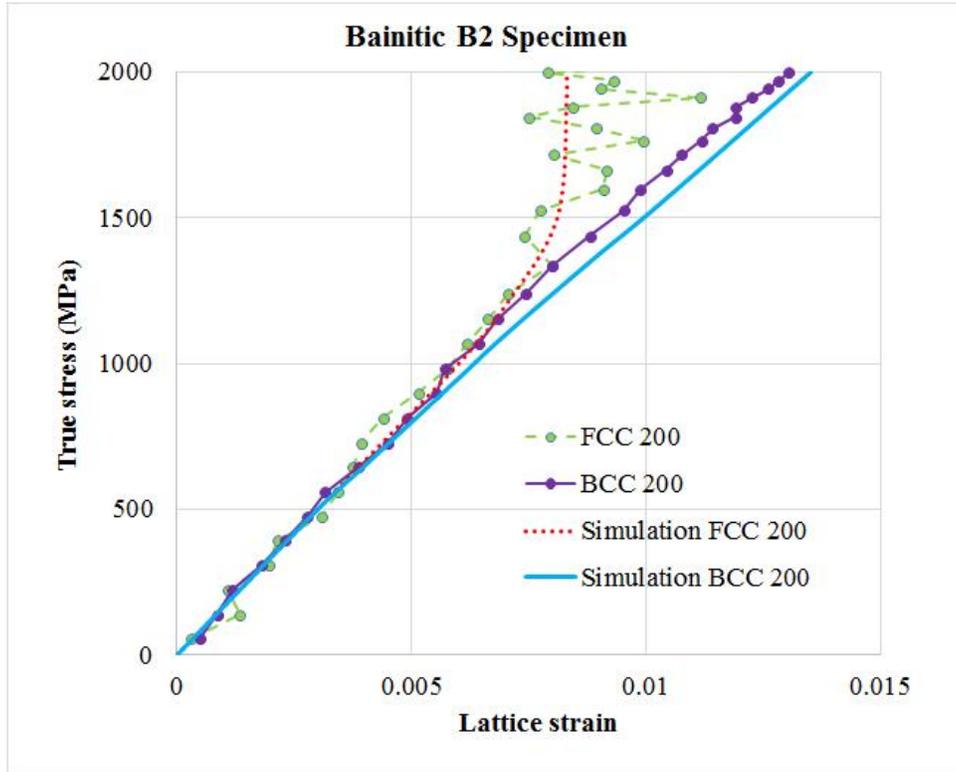

**Fig. 9:** CPFE Estimation and Experiment Results for lattice strains in Bainitic B2 Specimen

The modeling results using the material constants in Tables 3 and 4 match well with the empirical data collected using in situ neutron diffraction on the specimens, as shown in Fig. 7, 8 and 9. The material model and the corresponding parameters have been used to study and analyze the differences in the corresponding material parameters to understand the influence of the elastic, plastic and transformation plasticity components on the stability of the retained austenite within the martensitic and bainitic matrices. The lattice strain results for grains along the <200> direction were compared in Figures 7, 8 and 9 with the modeling results from the CPFE model. It can be seen that using the multiplicative decomposition of the deformation gradient and by determining the appropriate critical transformation driving force and hardening coefficients it is possible to capture the lattice strain response of the different microstructures. The austenite transformation is observed at true stress that is close to the macroscopic yield stress from Fig. 3 in all three samples. Moreover, we can also see that there is little change in the martensitic or bainitic phases for most part of the loading regime. The model is able to capture this well.

## 5. Discussion

Comparing the results from the neutron diffraction experiment and the crystal plasticity model we can see that while the elastic response is identical and the material constants are relatively same, the plastic deformation and the onset of transformation, however, differ greatly in all three cases. From Table 4, the initial drag stress $g_0^{(\alpha)}$ was slightly lower for the bainitic microstructures, which is expected as the martensite is a harder phase than the bainite. It can also be observed that the critical driving force for the transformation was estimated to be 21% higher for the martensitic microstructure compared to B1 — demonstrating that in the bainitic microstructure, the onset of transformation is much earlier, primarily because the threshold limit for the critical transformation driving force is lower. This observation is further validated by the fact that there is little difference in the strain accumulated within the BCC phases of the B1 and M1 samples. On the other hand, the transformation hardening coefficient $Q$ is 14% higher for B1 than for M1; showing that, in the bainitic microstructure after the onset of RA transformation the resistance to subsequent transformation is increasing, thereby causing the austenite to behave as a more stable volume. This can also be attributed to the factor that the retained austenite in the bainitic microstructures (B1 and B2) has a higher amount of carbon than the martensitic specimen (M1).

From the modeling material parameters in Table 4, we can also see that the major portion of the initial inelastic portion of the deformation is influenced by the transformation of the RA to martensite in both the B1 and M1 specimens, and that the surrounding microstructure directly affects the critical driving force and transformation hardening coefficients that determine the overall evolving stress state in these specimens. As a result, the modeling data and the material parameters estimated using this model confirm our hypothesis that the predominant factor affecting the stability of the RA in these specimens (B1 and M1) is the effect of the surrounding microstructure. The higher-strength martensite directly impacts the critical driving force necessary for the transformation, causing the transformation to occur at a slightly higher strain than in B1. In addition, we can observe from Table 4 that the micro-hardening parameters for M1 and B1 are identical in this case. So, the only parameter affecting the rate of transformation into martensite is the transformation hardening coefficient $Q$. As a result, we conclude that the difference in the rate of transformation is directly influenced by the microstructure morphology in B1 and M1. It is in line with our initial hypothesis that the primary reasons this transformation is more rapid in M1 is the lack of any film-like austenite and the potential of the blocky austenite in M1 to transform faster. This conclusion can be used to further infer that, once the austenite starts to strain plastically, the number of shear bands per unit volume of austenite is higher in the martensitic M1 than in the bainitic B1.

In order to further understand the effect of the bainitic microstructure and its influence on austenite stability, we compared specimens B1 and B2, both of which have a bainitic microstructure but different amounts of initial RA. From Figs. 3 and 4, we can clearly see that B2 with 9% initial RA has much more stable RA than B1 with 18% initial RA. This is expected, since the lower initial volume fraction of austenite limits the number of available shear bands for martensite nucleation [38]. The quantified micro-mechanical parameters for B1 and B2 directly demonstrate this effect — the critical driving stress for transformation in B2 is significantly higher than that in B1. However, it is interesting to note that the transformation hardening coefficient from Table 4 and the corresponding slope of the RA volume fraction from Fig. 3 remain relatively constant. This demonstrates that the majority of the RA transforming to martensite in the bainitic microstructure is the blocky RA. Because the amount of available blocky RA is smaller in B2, the transformation occurs only when the corresponding austenite is plastically strained. However, once plastic strain was induced, the observed transformation rate did not change significantly. Based upon this we can hypothesize that the number of shear bands available at the austenite lattice level is essentially identical with both B1 and B2 microstructures.

## 6. Conclusions

This work presents a first-of-its-kind effort to quantify the mechanical stability of retained austenite in martensitic and bainitic bearing steels by comparing the empirical response from in-situ neutron diffraction to the computational response from crystal plasticity modeling. The results have shown that the retained austenite in a martensitic microstructure has a higher critical driving force compared to that of a bainitic microstructure. This has been explained by the fact that the harder martensitic phase would partially shield the austenite from being plastically deformed under lower macroscopic loads. In addition, we found that once plastically deformed, a shear band intersection would generate a martensitic nucleation with higher probability in the austenite of the martensitic microstructure. To further understand the stability of retained austenite in a bainitic microstructure, the effect of the initial volume fraction of retained austenite was studied. The results show that, as expected, the higher volume fraction of initial RA results in an earlier RA transformation. However, if a large proportion of the RA transforming under plastic strain is blocky RA, the probability that a shear band will generate a martensitic nucleation that is relatively constant independent of the initial volume fraction of RA is increased. This allows the rate of transformation to be constant after the onset of RA transformation in the bainitic microstructures.

**Acknowledgments**

The authors thank Dr. Stephen P. Johnson (Director, Timken Technology Center) for permission and support of this work. We would also like to thank Mr. Matt Boyle, Mr. Chris Akey, Mr. Robert Pendergrass and the associates in the Prototype department for sample preparation.

A portion (neutron diffraction) of this research used resources at the Spallation Neutron Source, a DOE Office of Science User Facility operated by the Oak Ridge National Laboratory (ORNL).